\documentclass[journal]{vgtc}                     % final (journal style)
\onlineid{9332}

\vgtccategory{Research}

\title{\dsname{}: Evaluating and Collecting Human Sketches \\for MLLM-assisted Chart Annotation}

\author{%
  Yoonjae Oh\thanks{Both authors contributed equally to this research.},
  Seon Gyeom Kim\footnotemark[1],
  Jae Young Choi,
  Ryan Rossi,
  Jihyung Kil,
  Eunyee Koh,
  Tak Yeon Lee\thanks{Corresponding author.}
}

\authorfooter{
\item
Yoonjae Oh, Seon Gyeom Kim, and Tak Yeon Lee are with KAIST.
E-mails: \{angelaoh, ksg\_0320, takyeonlee\}@kaist.ac.kr.

\item
Jae Young Choi is with Texas A\&M University.
E-mail: jaeyoungchoi@tamu.edu.

\item
Ryan Rossi is with Patly.ai.
E-mail: ryan@patly.ai.

\item
Jihyung Kil and Eunyee Koh are with Adobe Research.
E-mails: \{jkil, eunyee\}@adobe.com.
}

\abstract{
    As multimodal large language models~(MLLMs) support a growing range of input modalities, increasing work explores how to incorporate rough sketches to convey user intent.
    For annotated chart generation, it remains unclear what annotation sketches people provide and when such visual input helps MLLMs generate more useful annotations.
    In this study, we examine when sketch input is useful for MLLM-generated chart annotations across variation in chart type and caption type.
    In addition, we qualitatively analyze participants' explanations of their output preferences to characterize what made generated annotations more or less helpful.
    To further document participants' annotation sketches, we present \dsname{}, comprising 1,600 annotation sketches collected across 160 chart-caption pairs from the conditions in which sketch guidance proved most beneficial, together with participants' annotation intents, perceived comprehension difficulty, and self-reported expressive limitations.
    We also label these sketches with structured metadata describing how each sketch relates to its caption and how participants express annotations through visual marks. 
    Together, our study and \dsname{} help determine when to solicit sketch input and provide empirical source for how people sketch chart annotations to support captions. The dataset and supplemental materials are available in our \href{https://osf.io/ad5r6/overview?view_only=791aee36e8b54cafa6a57c1cc25b6438}{OSF repository}.
}


\graphicspath{{figs/}{figures/}{pictures/}{images/}{./}} % where to search for the images

\usepackage{booktabs}
\usepackage{lipsum}
\usepackage{mwe}
\usepackage{ccicons}
\usepackage{listings}
\usepackage{tabularx}
\usepackage{array}
\AtEndPreamble{%
  \hypersetup{
    colorlinks=true,
    linkcolor=blue
  }%
}

\lstdefinestyle{judgeprompt}{
  basicstyle=\ttfamily\footnotesize,
  frame=single,
  framerule=0.4pt,
  framesep=6pt,
  xleftmargin=0.8em,
  xrightmargin=0.8em,
  breaklines=true,
  breakatwhitespace=true,
  breakindent=0pt,
  columns=fullflexible,
  keepspaces=true,
  showstringspaces=false,
  tabsize=2,
  captionpos=t,
  aboveskip=0.8\baselineskip,
  belowskip=1.2\baselineskip
}

\usepackage{mathptmx}
\usepackage{xspace}

\newcommand{\dsname}{AnnoSketch}
\newcommand{\mllmmodel}{gemini-3.1-flash}
\newcommand{\etal}{et~al.\@\xspace}

\begin{document}

%%%%%%%%%%%%%%%%%%%%%%%%%%%%%%%%%%%%%%%%%%%%%%%%%%%%%%%%%%%%%%%%
%%%%%%%%%%%%%%%%%%%%%% START OF THE PAPER %%%%%%%%%%%%%%%%%%%%%%
%%%%%%%%%%%%%%%%%%%%%%%%%%%%%%%%%%%%%%%%%%%%%%%%%%%%%%%%%%%%%%%%

% \firstsection{Introduction}

\maketitle

% \maketitle
\section{Introduction}
\label{sec:intro}

Captions and visual annotations have long been added to charts to help readers access specific data-related information.
Visual annotation emphasizes specific chart elements through graphical representations. 
This includes highlighting existing marks by adjusting visual encodings such as color, opacity, or size, and adding graphical elements such as arrows, brackets, and bounding boxes~\cite{ren2017chartaccent, rahman2025survey}. 
Such annotations provide an efficient means of directing readers’ attention to relevant information and can improve comprehension, memorability, and recall~\cite{bateman2010useful, borkin2015beyond, chun2020giving}.
Captions allow authors to convey domain or analytic knowledge related to the chart in free-form text~\cite{lundgard2021accessible}.
A caption may describe an insight that is not visually prominent in the chart.
However, when visually salient information aligns with what the caption describes, readers are more likely to take away that information than other insights from the chart~\cite{kim2021towards}.
This suggests that visual annotation can support caption understanding by making the insight described in the caption stand out, even when that insight is less salient in the original chart.

\begin{figure*}[t]
    \centering
    \includegraphics[width=\textwidth]{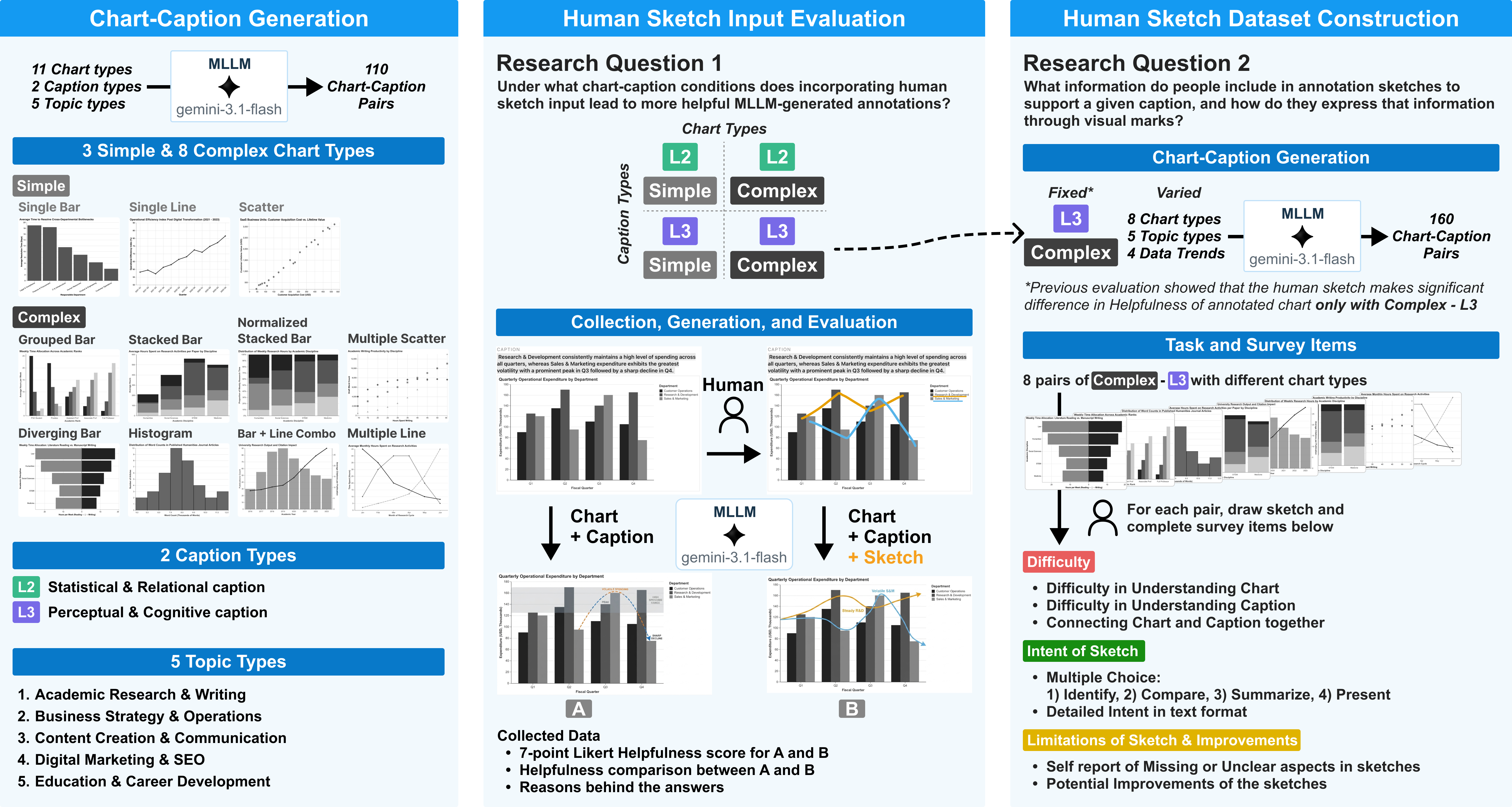}
    \caption{
    Overview of our two-stage study.
    We first construct a chart-caption corpus spanning diverse chart types, caption levels, and topics~(left).
    We then evaluate when conditioning an MLLM on human sketches improves generated visual annotations across chart~complexity and caption~level~(middle).
    Based on this result, we focus the subsequent data collection on Complex-L3 pairs to characterize what information people include in annotation sketches and how they express it through visual marks~(right).
    }
    \label{fig:overview}
    % \vspace{-8pt}
\end{figure*}

Annotated charts are widely used in practice, and prior studies have characterized the functions and visual designs of annotations in real-world examples. 
Prior work has also proposed structural representations of annotations that are compatible with well-established visualization grammars such as Vega~\cite{chen2025chartmark}.
However, these representations assume familiarity with visualization structures and annotation grammars, making them unsuitable as low-effort input mechanisms for non-expert authors.
While this work provides formal accounts of annotations, less is known about how non-expert authors communicate annotation ideas through informal forms of input, such as natural-language descriptions and rough sketches. 
Recent multimodal large language models~(MLLMs) can incorporate such informal input into visual generation, potentially enabling non-experts to turn preliminary annotation ideas into more polished outputs. 
Conditioning such generation on human input, however, carries known risks: freehand sketches are inherently abstract and distorted, and existing conditioning mechanisms often fail to capture what they intend~\cite{bourouis2026sketchingreality}; models can also favor responses that match what a user supplied over more accurate ones~\cite{sharma2024towards}.
For annotation generation in particular, we still lack an understanding of what kinds of low-effort rough human input can be useful for AI systems, and when such input is worth asking for at all. 
While prior work has characterized completed annotations in existing charts, less is known about what information people choose to express when sketching an annotation idea for a given chart and caption, and how they represent that information through rough visual marks.
From this perspective, we consider the following research questions:
\textbf{RQ1.} Under what chart-caption conditions does incorporating human sketch input lead to more helpful MLLM-generated annotations?
\textbf{RQ2.} What information do people include in annotation sketches to support a given caption, and how do they express that information through visual marks?
To address these questions, we conducted a two-stage study~(Fig.~\ref{fig:overview}).
We first evaluate when human sketch input improves MLLM-generated annotations across different chart-caption conditions, and then use the findings from this evaluation to guide the construction of a human sketch dataset for further analysis.

To answer \textbf{RQ1}, we constructed a synthetically generated chart-caption corpus spanning multiple chart types and topics, while varying two experimental dimensions: \emph{chart composition}~(simple vs.\ complex) and the \emph{semantic level of the caption}~(L2,~statistical and relational, vs.\ L3,~perceptual and cognitive~\cite{lundgard2021accessible}), yielding four conditions.
In a crowdsourced study, participants first sketched their annotation ideas directly on the chart to support the given caption.
They then evaluated and compared two annotated charts generated by an MLLM: one from the chart and caption alone, and the other additionally conditioned on the participant's sketch. Participants were not informed which input condition produced each output.
We found that the sketch-conditioned output was preferred significantly more often only for complex charts paired with L3 captions. 
For the other three conditions, sketch input produced no measurable advantage~(\S\ref{sec:sketchinput}).

Because the first study suggested that human sketch input was particularly valuable for complex charts paired with L3 captions, we focused our dataset construction on this condition to answer \textbf{RQ2}. 
To broaden the empirical coverage of the collection, we expanded the chart-caption corpus to include diverse underlying data patterns and relationships. 
We then built \dsname{}~(\S\ref{sec:dataset}), a collection of 1,600 human-drawn annotation sketches across 160 chart-caption pairs, with ten independently drawn sketches for each pair.
Participants also described their annotation intent, what they found difficult to express through drawing, and their perceived difficulty in understanding and relating the chart and caption.
We further developed structured metadata that characterizes how each sketch relates to the caption and chart and how individual marks serve different communicative functions. 
Using this metadata, we analyze patterns in caption coverage and extension, as well as relationships between annotation forms and roles.
We found recurring but non-exclusive associations between visual forms and communicative roles, with the same visual form serving multiple functions across the collection. 
This pattern suggests that the communicative role of a rough mark cannot be inferred from its visual form alone.

Our key contributions are:
\begin{itemize}
  \item We identify the chart-caption conditions under which incorporating rough human visual input leads users to prefer MLLM-generated annotations, as well as conditions in which it provides no measurable preference advantage, through a 2$\times$2 crowdsourced study of chart composition and caption semantic level.
  \item We release \dsname{}, a collection of 1,600 human-drawn annotation sketches across 160 Complex–L3 chart–caption pairs, with ten independent sketches per pair and accompanying participant reports on annotation intent, expressive limitations, and perceived difficulty.
  \item We develop a structured metadata schema that characterizes how sketches relate to their chart and caption and how individual marks serve communicative functions. We apply the schema to the full collection and validate the resulting labels against human annotations.
\end{itemize}
\section{Related Work}
\label{sec:rw}

Our work lies at the intersection of visual annotation for data visualization, human visual guidance for multimodal generation, and datasets for chart understanding and annotation. 
We first review how prior work has characterized, authored, and generated graphical overlays for charts. We then discuss lightweight visual input as a means of communicating human intent to generative models. Finally, we position our dataset against existing chart and sketch datasets.

\subsection{Graphical Overlays for Data Visualization}
\label{sec:related:overlays}

Visual annotations augment charts with graphical or textual cues that direct attention, clarify relationships, and emphasize interpretations of the data. 
Kong and Agrawala~\cite{kong2012graphical} introduced \emph{graphical overlays} as visual elements layered over a chart to support perceptual and cognitive processes during chart reading. 
Subsequent work characterized the design and communicative roles of such cues. 
For example, Ren~\etal~\cite{ren2017chartaccent} derived a design space from annotated news graphics, distinguishing annotation forms and the chart components to which they are anchored. 
Rahman~\etal~\cite{rahman2024qualitative} developed a taxonomy of common annotation types and their roles.
Kong~\etal~\cite{kong2017internal} distinguished \emph{internal cues}, such as color or transparency changes to existing elements, from \emph{external cues}, such as arrows or outlines added to the chart.
This distinction is particularly relevant to our work, as rough human drawings may express annotation intent by either modifying the appearance of existing chart elements or introducing new marks around them.

Beyond characterizing annotation designs, prior work has investigated how annotations can be authored and generated, with a gradual shift of effort from the author to the system.
At one end of this spectrum, authoring tools keep the author in full control.
These systems enable authors to interactively place data-driven annotations~\cite{ren2017chartaccent}, define annotations through structured grammars~\cite{chen2025chartmark}, or formalize freeform sketches drawn directly on charts~\cite{lin2023inksight}.
In contrast, automated approaches derive annotations from accompanying text by linking described information to relevant chart elements or underlying data~\cite{kong2014extracting,lai2020automatic,hullman2013contextifier,hao2024finflier}.
ChartAnno~\cite{chen2026chartanno} further demonstrated this dependence, showing that MLLM-generated annotations degrade especially when an instruction does not clearly convey the author's intent.

Prior work has largely focused on analyzing completed annotations, and supporting their specification and generation through structured or interactive methods. 
We instead examine an earlier stage of the authoring process: we ask how non-expert authors express annotation intent through rough drawings and when such input can inform generation.

\subsection{Human Visual Guidance for Multimodal Generation}
\label{sec:related:humanguidance}

Natural language provides an intuitive way to convey semantic intent to generative models, but it is less effective for specifying spatial constraints, such as \emph{where} content should appear or \emph{which} visual elements should be modified. 
Controllable generation methods have therefore supplemented language with visual conditions that encode spatial or structural information. 
For example, ControlNet~\cite{zhang2023adding} conditions pretrained diffusion models on representations such as edges, depth maps, segmentation maps, and human poses, and related methods condition generation on bounding-box layouts or auxiliary structural adapters~\cite{li2023gligen, mou2024t2iadapter}.
These methods demonstrate the value of visual information for controlling generation, but typically rely on structured spatial conditions such as edge, depth, segmentation, or pose representations, rather than informal expressions of human intent.

More recent work has explored visual inputs that can be expressed more directly
by users. 
Set-of-Mark~\cite{yang2023som}, for example, demonstrates that structured, machine-generated marks over image regions can substantially improve visual grounding in multimodal models.
ViP-LLaVA~\cite{cai2024vipllava} enables multimodal models to interpret visual prompts such as points, bounding boxes, and arrows drawn directly on images so that users can reference regions without specifying their locations in text prompts. 
MagicQuill~\cite{liu2025magicquill} further supports image editing through lightweight strokes, using an MLLM to infer editing intent from these interactions. 
These approaches move beyond machine-generated structured conditions toward visual input that resembles how people naturally indicate objects, regions, and intended changes.

However, lightweight human input also introduces ambiguity. 
Freehand sketches can abstract away visual details and contain shape distortions or spatial imprecision.
Koley~\etal~\cite{koley2024all} found that conditioning designed around precise, edge-map-like input struggles with abstract sketches and can propagate sketch deformities into generated outputs, motivating an abstraction-aware approach.
More recently, SketchingReality~\cite{bourouis2026sketchingreality} treats freehand scene sketches as a challenging input modality, emphasizing the need to recover semantic intent while tolerating deviations from precise image geometry. 
These findings suggest that rough drawings should not be treated as spatial constraints to be faithfully reproduced.  
Instead, they provide approximate and incomplete cues about what users consider important, which models may over-rely on even when those cues misrepresent the intended message~\cite{sharma2024towards}.

Rather than asking how faithfully a model can follow rough human visual guidance, we ask when it provides useful information beyond what an off-the-shelf MLLM can infer from the chart and caption alone.

\subsection{Datasets for Chart Understanding and Annotation}
\label{sec:related:dataset}
A growing body of datasets has supported the development and evaluation of machine chart understanding across tasks such as question answering, reasoning, and caption generation. 
Early chart question-answering datasets such as FigureQA~\cite{kahou2018figureqa}, DVQA~\cite{kafle2018dvqa}, and LEAF-QA~\cite{chaudhry2020leafqa} relied on synthetic charts and template-based questions, while ChartQA~\cite{masry2022chartqa} introduced real-world charts paired with human-written questions requiring visual and logical reasoning. 
More recent benchmarks evaluate MLLMs with broader chart diversity and more demanding reasoning, including ChartX~\cite{xia2024chartx}, ChartBench~\cite{xu2023chartbench}, and CharXiv~\cite{wang2024charxiv}, which draws charts from scientific papers. 
In parallel, chart captioning datasets pair charts with natural-language descriptions: SciCap~\cite{hsu2021scicap} collects figure-caption pairs from arXiv papers, Chart-to-Text~\cite{kantharaj2022chart} provides large-scale chart summarization benchmarks, and VisText~\cite{tang2023vistext} offers crowdsourced captions stratified by the semantic levels of Lundgard and Satyanarayan, the same framework we adopt to define our caption conditions. 
Across these datasets, text is the modality being produced or evaluated against a chart; none records the graphical marks that people layer \emph{onto} charts to communicate about them.

Freehand drawing itself has been captured at scale. 
Object-level collections such as TU-Berlin~\cite{eitz2012humans}, Sketchy~\cite{sangkloy2016sketchy}, and QuickDraw~\cite{ha2018neural} contain sketches of isolated objects, and FS-COCO~\cite{chowdhury2022fscoco} extends this line to freehand scene sketches drawn by non-expert individuals and paired with photographs and text descriptions. 
In these datasets, however, a sketch \emph{depicts} an object or scene, and the drawer's intent is recorded at most as a category label or a caption of the depicted content. 
They do not treat sketches as communicative overlays anchored to an underlying artifact, nor do they capture the gap between what a drawer meant to convey and what their strokes actually show.

Documenting that gap requires sketches paired with ground-truth of their authors' intent.
ChartAnno~\cite{chen2026chartanno} comes closest to our setting but evaluates the annotations that \emph{models} generate rather than documenting how \emph{humans} express annotation intent.
In \dsname{}, each sketch is anchored to a specific chart-caption pair and accompanied by the author's stated intent, self-reported expressive limitations, and perceived difficulty.
The collection enables empirical study of how non-expert authors translate intent into rough marks, where that translation breaks down, and how generation systems can bridge the difference.
\begin{figure*}[t]
    \centering
    \begin{subfigure}[t]{0.48\textwidth}
        \includegraphics[width=\linewidth]{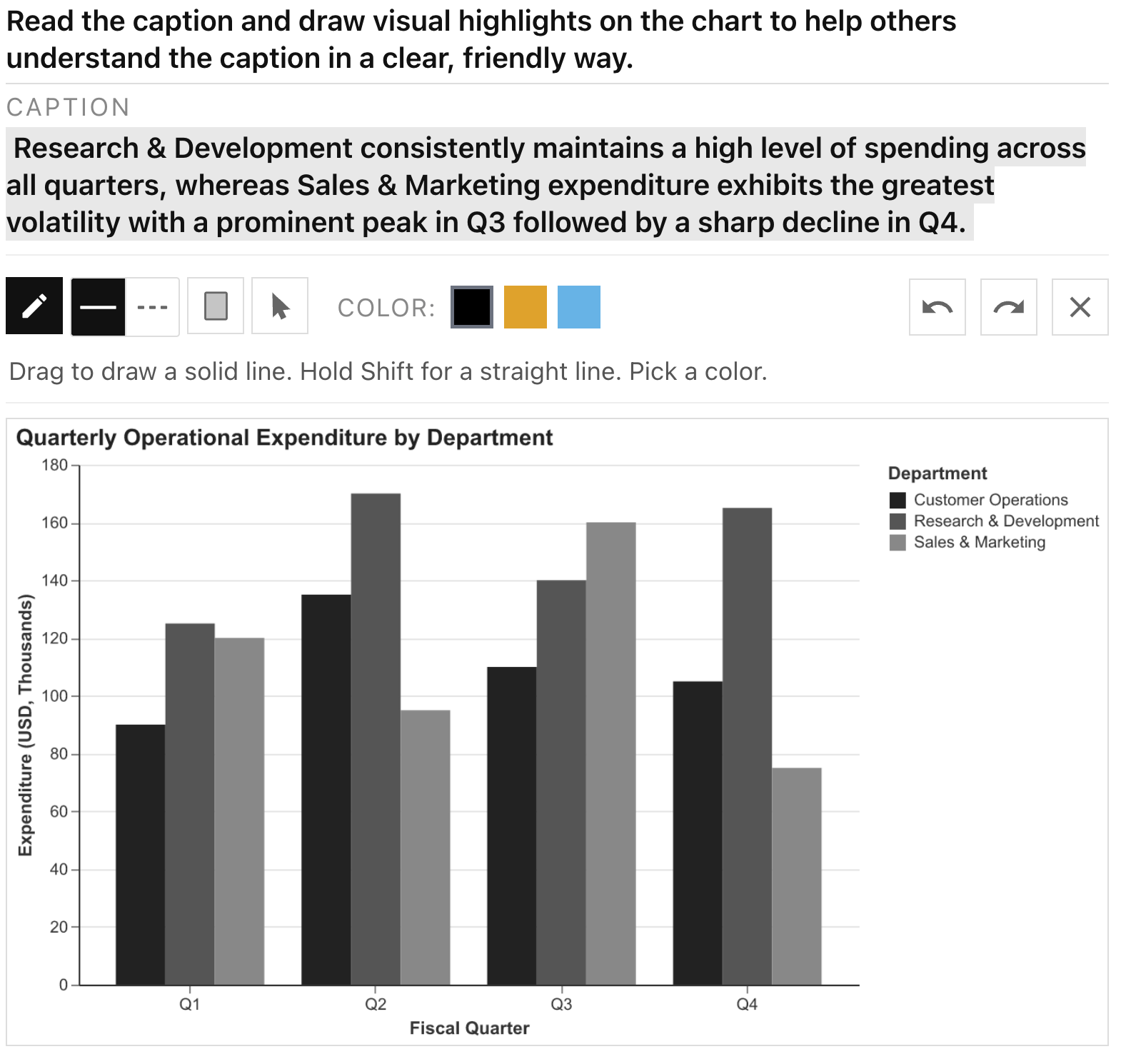}
        \caption{}\label{fig:drawing-interface}
    \end{subfigure}
    \hfill
    \begin{subfigure}[t]{0.48\textwidth}
        \includegraphics[width=\linewidth]{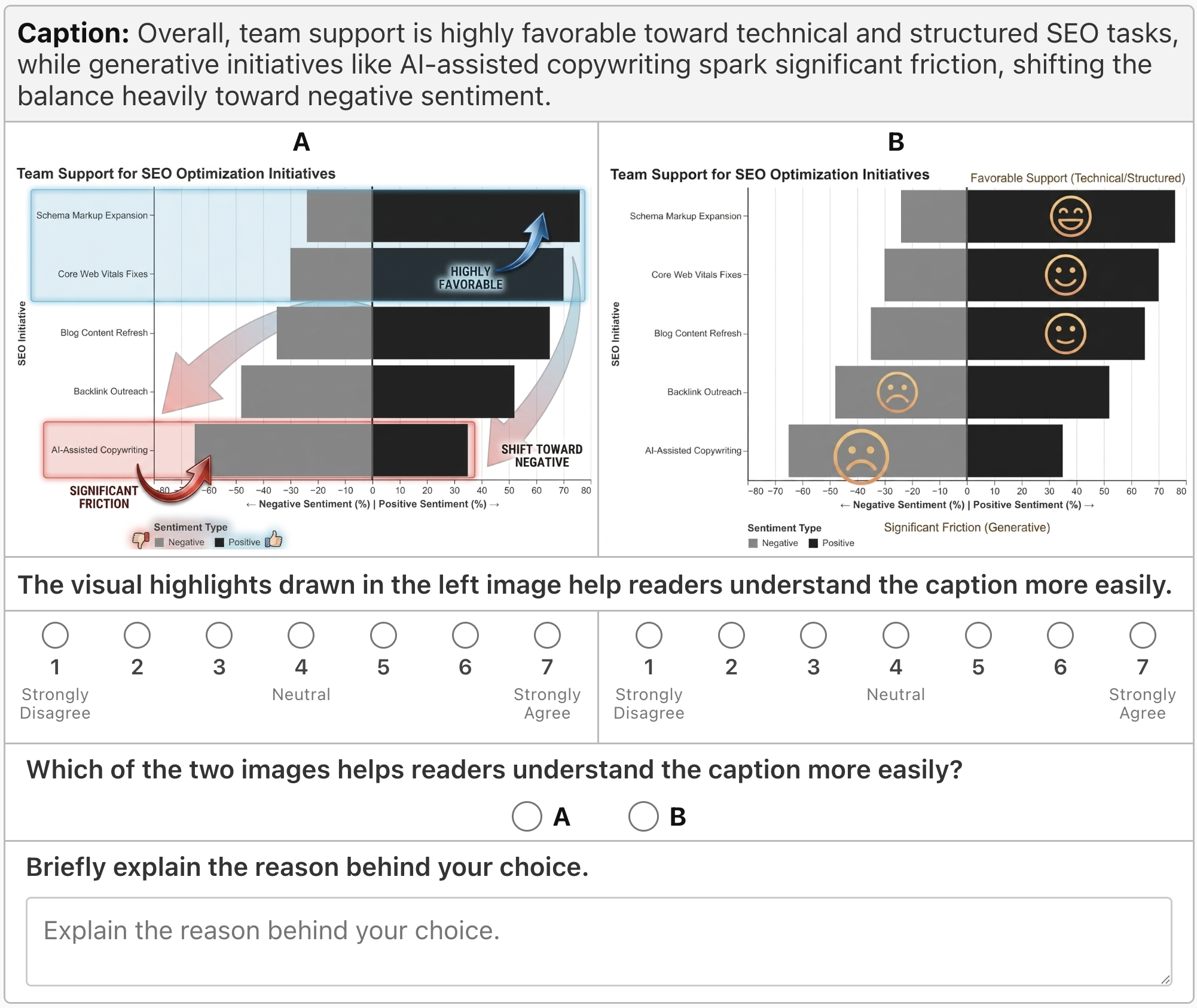}
        \caption{}\label{fig:evaluation-interface}
    \end{subfigure}
    \vspace{-6pt}
    \caption{
        Study interfaces. (a) Participants read a caption and drew visual highlights directly on the chart, using freehand drawing, solid and dashed lines, semi-transparent rectangles, and three color options. (b) Participants then viewed two annotated charts for the same caption, rated each on a seven-point Likert scale, selected the more helpful output, and briefly explained their choice.
    }
    \label{fig:interfaces}
    % \vspace{-8pt}
\end{figure*}

\section{Human Sketch Input Evaluation}
\label{sec:sketchinput}

We conducted a study to examine when human-provided sketches improve MLLM-generated visual annotations for supporting caption understanding.
Section~\ref{sec:sketchinput:study-design} introduces the experimental factors and comparison conditions, Section~\ref{sec:sketchinput:experiment-setup} describes the study materials and procedure, Section~\ref{sec:sketchinput:analysis} describes how we analyzed the ratings and preferences and how we coded participants' explanations of their choices, and Section~\ref{sec:sketchinput:study-results} reports both sets of results.

\subsection{Study Design}
\label{sec:sketchinput:study-design}
Prior work has shown that what people attend to in a visualization changes with the question they are trying to answer~\cite{wang2024salchartqa}.
This suggests that what should be emphasized may depend on how the caption relates to the chart.
We therefore considered whether the value of human sketch guidance might depend on how clearly the relevant marks or relationships could be identified from the chart and caption alone.
Accordingly, we varied both the kind of information conveyed by the caption and the composition of the chart.
First, following the model of Lundgard and Satyanarayan~\cite{lundgard2021accessible}, we considered L2 and L3 captions. 
While L2 captions describe statistical relations, such as extrema and comparisons linked to specific marks, L3 captions tend to capture broader perceptual patterns, such as trends and exceptions distributed across multiple marks or regions. 
Second, we considered multiple chart types and grouped them into simple and complex based on the number of series and the compositional structure of the chart, rather than on chart genre alone.
The simple charts were a single-series bar chart, line chart, and scatter plot.
The complex charts included a grouped bar chart, a stacked bar chart, a normalized stacked bar chart, a histogram, a diverging bar chart, a multiple line chart, a multiple scatter plot, and a bar and line combination chart. 
Complex charts contain more marks, series, and regions, and often more bins or categories, increasing the amount of visual information viewers may need to consider.
Prior work has found that reproduction error increases as the number of displayed values increases~\cite{mccoleman2021rethinking}.
This motivated us to include chart complexity as a factor that could affect the value of sketch guidance, particularly when more candidate elements or relationships are available for annotation.

Across these factors, we created a set of 110 chart-caption pairs by combining 11 chart types, five topics, and two caption levels. 
The five topics were Business Strategy \& Operations, Education \& Career Development, Academic Research \& Writing, Content Creation \& Communication, and Digital Marketing \& SEO. 
Each chart type was instantiated for all five topics and paired with an L2 caption and an L3 caption, producing 30 simple and 80 complex chart-caption pairs.

Both the charts and captions were generated using LLMs. 
For chart generation, we provided the model with a target chart type and topic and instructed it to produce a complete Vega-Lite specification~\cite{satyanarayan2016vega}. 
A shared generation guide specified a monochrome color palette, readable titles and axis labels, legends for charts with multiple series, and encoding requirements appropriate to each chart type. 
The model was also instructed to generate plausible data with varied patterns and to avoid overlapping or clipped visual elements. 
For caption generation, we provided the model with the rendered chart, chart metadata, and underlying data values, and instructed it to generate one L2 caption describing statistical or relational information and one L3 caption describing higher-level perceptual patterns.

\subsection{Experiment Setup}
\label{sec:sketchinput:experiment-setup}
Using this stimulus set, we recruited 60 participants via CloudResearch\footnote{\url{https://www.cloudresearch.com/}}. All participants were proficient in English and had normal or corrected-to-normal vision. Participants ranged in age from 24 to 63 years ($M = 38.7$, $SD = 9.7$), including 37 men (61.7\%), 22 women (36.7\%), and one non-binary participant (1.7\%). Each participant received USD 8 for their participation, and the study took approximately 45 minutes on average.
Before the main study, participants completed a tutorial by tracing an example annotation.
Each participant was then assigned to one of 20 predefined task sets and evaluated six chart-caption pairs in a randomized order: one Simple-L2, one Simple-L3, and two Complex-L2 and two Complex-L3 pairs in alternation. Three participants completed each set.
For each pair, participants read the caption and drew visual highlights directly on the corresponding chart using the interface shown in Fig.~\ref{fig:drawing-interface}. 
The interface supported freehand drawing, solid and dashed lines, semi-transparent rectangles, three color options, and a selection tool for editing existing annotations.

For each pair, we pre-generated a baseline annotation using only the original chart and caption. 
This baseline was shown to all participants who evaluated the same chart–caption pair.
For each completed drawing, we generated a sketch-conditioned annotation using the original chart, caption, and participant's drawing. 
Both outputs were generated using \mllmmodel. 
The generation prompt instructed the model to preserve the original chart while adding legible visual annotations that emphasized the chart elements and relationships relevant to the caption.
For the sketch-conditioned condition, the model was additionally instructed to infer the participant's annotation intent from the rough drawing and refine it rather than directly reproducing the sketch.
The model drew the visual annotations directly over the original chart image, and no additional rendering or post-processing was applied. 
After drawing annotations on six charts, participants reviewed the baseline and sketch-conditioned outputs for each pair using the interface shown in Fig.~\ref{fig:evaluation-interface}.
The outputs were labeled A and B without revealing their generation conditions, and the assignment of the outputs to the two positions was randomized.
Participants rated how well each output supported caption understanding on a seven-point Likert scale, selected the more helpful output, and briefly explained their choice.

\subsection{Analysis}
\label{sec:sketchinput:analysis}

For each condition, we compared the helpfulness ratings of the two outputs with paired Wilcoxon signed-rank tests, and tested the proportion of comparisons favoring the sketch-conditioned output against chance with binomial tests.
We report Holm-adjusted p-values across the four conditions.
We also tested each output type across conditions with paired Wilcoxon signed-rank tests on participant-level means for the two design factors, chart complexity and caption level, to locate the source of the difference.
This allowed us to determine whether effects arose from the baseline, sketch-conditioned output, or both.

To examine the reasons underlying participants' preferences between the two generated images, we conducted a qualitative analysis of all 360 open-ended responses explaining their choices, alongside the corresponding sketches, baseline images, and sketch-conditioned images.
Two researchers first independently open-coded a random sample of 100 responses and iteratively refined the resulting codes through discussion by consolidating recurring concepts, merging overlapping categories, and clarifying ambiguous code boundaries~\cite{saldana2009coding}.
This process produced a codebook of shortcomings observed in non-preferred images.
Across both baseline and sketch-conditioned images, common shortcomings included \textit{Cluttered and Noisy} and \textit{Inaccurate Generation}.
Baseline-specific shortcomings included \textit{Weak or Ambiguous Encoding}, \textit{Lack of Key Information}, and \textit{Misleading or Biased Framing}, whereas \textit{Altering Original Chart} was specific to sketch-conditioned images.

Separately, for each comparison, we examined whether the sketch-conditioned output visually corresponded to the sketch.
We coded correspondence as present when the generated annotations retained at least part of the sketch's graphical characteristics, such as approximate location, color, or annotation form, while using those characteristics for an annotation function consistent with the corresponding part of the sketch.
Exact spatial overlap or identical annotation form was not required, and moderate spatial displacement was allowed.
Graphical similarity alone was not considered sufficient when the corresponding feature served a different annotation function in the output.
These judgments concerned the visual relationship between the sketch and generated annotations and were made independently of whether either expressed information stated in the caption.
For comparisons in which the sketch-conditioned output was preferred, we additionally identified annotation forms added beyond the original sketch using five categories: \textit{Text}, \textit{Line}, \textit{Color}, \textit{Enclosure}, and \textit{Glyph}, adapted from prior work~\cite{rahman2024qualitative}.

The two researchers independently applied the coding scheme to all 360 comparisons.
Inter-rater reliability was assessed using Cohen's $\kappa$, treating each multi-label code as a binary present-or-absent variable~\cite{10.1145/3359174}. 
Mean $\kappa$ was .86 (range = .69--.95) for baseline shortcomings, .74 (range = .67--.82) for sketch-conditioned shortcomings, and .87 (range = .77--.95) for additional annotation forms. 
For sketch-reflection judgments, $\kappa$ was .72. 
All remaining discrepancies were resolved through discussion to reach consensus.

\subsection{Results}
\label{sec:sketchinput:study-results}
Human-provided sketches improved the generated annotations only where the model performed poorly on its own.
Baseline quality varied with the conditions while sketch-conditioned quality did not.
Baseline ratings were lower for complex than simple charts ($M=4.93$ vs.\ $5.72$, $p<.001$) and for L3 than L2 captions ($M=4.92$ vs.\ $5.47$, $p<.01$), whereas ratings of the sketch-conditioned outputs differed neither by chart complexity ($M=5.47$ vs.\ $5.66$, $p=.14$) nor by caption level ($M=5.51$ vs.\ $5.56$, $p=.68$).
The two sources of difficulty accumulate in Complex-L3, where the unaided model performed worst, and it was there that participants preferred the sketch-conditioned output.
In this condition, sketch-conditioned outputs received higher ratings ($M=5.57$, $SD=1.49$) than baseline outputs ($M=4.56$, $SD=1.85$), with a mean difference of $1.01$ points (Wilcoxon signed-rank test, $p<.01$).
Participants preferred them in 79 of 120 comparisons (65.8\%), which was significantly higher than chance (binomial test, $p<.01$).
We found no significant rating differences between the baseline and sketch-conditioned outputs in the Simple-L2, Simple-L3, or Complex-L2 conditions (all n.s.). 
Direct preferences for the sketch-conditioned outputs were $40.0\%$, $46.7\%$, and $50.0\%$, respectively, and none differed significantly from chance (all n.s.). 
Sketch preference showed no consistent advantage for either output in the Complex-L2 tasks while exceeding $65\%$ in both Complex-L3 tasks.
These results suggest that sketch input was most useful when complex charts were paired with captions describing broader perceptual patterns.

To better understand what made individual outputs preferable, we next examined participants' open-ended explanations alongside the corresponding sketches and generated images across all 360 comparisons (Table~\ref{tab:qualitative_results}). 
Participants preferred the sketch-conditioned image in 191 cases (53.1\%) and the baseline image in 169 cases (46.9\%). 
Notably, even when the baseline image was preferred, the sketch-conditioned output still visually corresponded to the participant's sketch in 129 of 169 cases (76.3\%), suggesting that visual correspondence with the sketch alone did not necessarily lead to preference.
Among the preferred sketch-conditioned outputs, 173 of 191 (90.6\%) visually corresponded to the participant's sketch, while 121 (63.4\%) introduced at least one additional annotation form beyond those explicitly present in the original sketch.
Text was the most common addition (97, 50.8\%), followed by lines (43, 22.5\%), enclosures (37, 19.4\%), color (29, 15.2\%), and glyphs (22, 11.5\%).
Together, these results suggest that preferred sketch-conditioned outputs tended not only to maintain visual correspondence with participants' sketches but also to extend them with additional visual cues beyond what was explicitly sketched.

Participants' explanations for non-preferred outputs revealed both shared and condition-specific shortcomings. When the baseline image was rejected, the most frequent issue was \textit{Weak or Ambiguous Encoding} (16, 8.4\%), followed by \textit{Cluttered and Noisy} (15, 7.9\%), \textit{Inaccurate Generation} (11, 5.8\%), \textit{Lack of Key Information} (10, 5.2\%), and \textit{Misleading or Biased Framing} (7, 3.7\%). When the sketch-conditioned image was rejected, \textit{Inaccurate Generation} (20, 11.8\%) and \textit{Altering Original Chart} (14, 8.3\%) were the dominant issues, with \textit{Cluttered and Noisy} appearing in ten cases (5.9\%). While both approaches were undermined by inaccurate or cluttered marks, baseline outputs were additionally prone to missing, ambiguous, or misleading information.

\begin{table}[t]
    \centering
    \caption{Helpfulness ratings and preferences for baseline and sketch-conditioned annotations across conditions. 
    Ratings were measured on a seven-point Likert scale and are reported as mean (SD). 
    The $p$-values are from paired Wilcoxon signed-rank tests, Holm-adjusted across the four conditions.
    Sketch preference is the percentage of comparisons in which the sketch-conditioned output was selected as more helpful.}
    \label{tab:sketch-results}
    \scriptsize
    \setlength{\tabcolsep}{3pt}
    \begin{tabular}{lcccc}
        \toprule
        Condition &
        w/o Sketch &
        w/ Sketch &
        Rating $p$ &
        Sketch pref. \\
        \midrule
        Simple-L2
        & 5.80 (1.64)
        & 5.78 (1.42)
        & n.s.
        & 40.0\% \\

        Simple-L3
        & 5.63 (1.37)
        & 5.53 (1.61)
        & n.s.
        & 46.7\% \\

        Complex-L2
        & 5.30 (1.52)
        & 5.37 (1.65)
        & n.s.
        & 50.0\% \\

        Complex-L3
        & \textbf{4.56 (1.85)}
        & \textbf{5.57 (1.49)}
        & $\mathbf{<.01}$
        & \textbf{65.8\%} \\
        \bottomrule
    \end{tabular}
    % \vspace{-8pt}
\end{table}

\begin{table}[t]
\centering
\caption{Qualitative coding results for image preferences, including reasons for not preferring each condition, sketch reflection by the sketch-conditioned image, and forms of annotation added to the sketch.}
\label{tab:qualitative_results}
% \vspace{-4pt}

\scriptsize
\renewcommand{\arraystretch}{0.92}
\setlength{\tabcolsep}{3pt}

\begin{tabularx}{\columnwidth}{
    @{}
    >{\raggedright\arraybackslash}X
    >{\raggedleft\arraybackslash}p{0.28\columnwidth}
    @{}
}
\toprule

\multicolumn{2}{@{}l}{\textbf{Preference} ($N=360$)} \\
\cmidrule{1-2}
Baseline preferred
    & 169 (46.9\%) \\
Sketch-conditioned preferred
    & 191 (53.1\%) \\

\midrule

\multicolumn{2}{@{}l}{\textbf{Baseline}} \\
\midrule

\multicolumn{2}{@{}l}{
    \textbf{\textit{Reasons for Not Preferring Baseline}} ($n=191$)
} \\
Weak or Ambiguous Encoding & 16 (8.4\%) \\
Cluttered and Noisy & 15 (7.9\%) \\
Inaccurate Generation & 11 (5.8\%) \\
Lack of Key Information & 10 (5.2\%) \\
Misleading or Biased Framing & 7 (3.7\%) \\

\midrule

\multicolumn{2}{@{}l}{\textbf{Sketch-conditioned}} \\
\midrule

\multicolumn{2}{@{}l}{
    \textbf{\textit{Reasons for Not Preferring Sketch-conditioned}} ($n=169$)
} \\
Inaccurate Generation & 20 (11.8\%) \\
Altering Original Chart & 14 (8.3\%) \\
Cluttered and Noisy & 10 (5.9\%) \\

\addlinespace[4pt]

\multicolumn{2}{@{}l}{
    \textbf{\textit{Whether the Sketch-conditioned Image Reflected the Sketch}}
} \\
Baseline preferred ($n=169$)
    & 129 (76.3\%) \\
Sketch-conditioned preferred ($n=191$)
    & 173 (90.6\%) \\

\addlinespace[4pt]

\multicolumn{2}{@{}l}{
    \textbf{\textit{Forms of Annotation Added to the Sketch}} ($n=191$)
} \\
Text & 97 (50.8\%) \\
Line & 43 (22.5\%) \\
Enclosure & 37 (19.4\%) \\
Color & 29 (15.2\%) \\
Glyph & 22 (11.5\%) \\

\addlinespace[2pt]
Responses with added form(s)
    & 121 (63.4\%) \\

\bottomrule
\end{tabularx}
% \vspace{-8pt}
\end{table}
\section{\dsname{}} 
\label{sec:dataset}

\begin{figure*}[t]
    \centering
    \includegraphics[width=\textwidth]{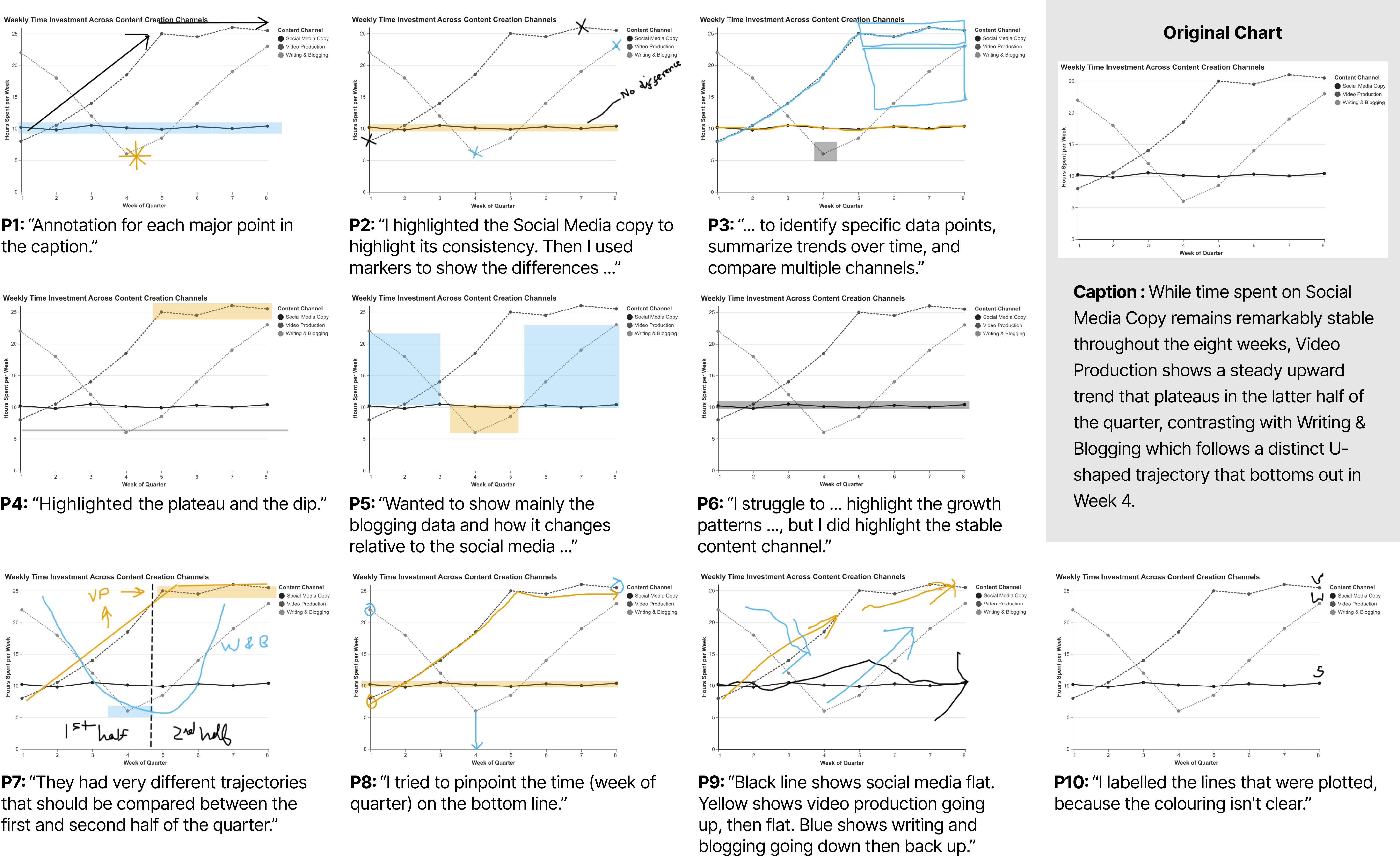}
    \caption{Ten annotation sketches drawn by different participants for the same chart-caption
    pair. 
    The sketches show differences and similarities in what participants emphasized and how they expressed it through visual marks. 
    The original chart-caption pair is shown at the top right, and each sketch is accompanied by the participant's description of their intent.}
    \label{fig:example-sketches}
    % \vspace{-8pt}
\end{figure*}

Building on the finding that sketch input was most beneficial for Complex-L3 chart-caption pairs~(\S~\ref{sec:sketchinput}), we construct \dsname{}, a dataset of human-drawn visual annotation sketches for this condition.
Our goal is to document what people choose to annotate and how they express annotation intent through rough sketches, thereby providing an empirical basis for future sketch-conditioned chart annotation systems.
\dsname{} comprises 1,600 annotation sketches, each accompanied by survey responses and structured metadata on how the sketch relates to its caption and how annotations are expressed through sketch marks.
Specifically, we created Complex-L3 chart-caption pairs and collected annotation sketches~(\S\ref{sec:dataset:collection}), developed a metadata schema with category definitions and illustrative examples~(\S\ref{sec:dataset:metadata_schema}), labeled all sketches with multiple LLMs and validated the labels against human annotations~(\S\ref{sec:dataset:labeling}), and analyzed the resulting quantitative patterns~(\S\ref{sec:dataset:annotation_patterns}).

\subsection{Data Collection}
\label{sec:dataset:collection}

We used the same eight complex chart types as in the first study: grouped bar, stacked bar, normalized stacked bar, histogram, diverging bar, multiple line, multiple scatter, and bar and line combination charts.
While fixing the caption level to L3 and restricting the stimuli to these complex chart types, we generated four chart-caption pairs per chart type and topic using the same \mllmmodel{} and overall generation procedure as in the first study, yielding 160 unique pairs.
In each iteration, we instructed the model to produce a data trend distinct from those already generated, which broadened the variation in underlying data patterns across the corpus.
To distribute these stimulus variations, we prepared 20 predefined task sets, each containing one chart-caption pair for each of the eight chart types, with topics and trend variations spread across the sets.
Together, the 20 sets comprised 160 chart-caption pairs.
We recruited 200 participants through CloudResearch. All participants were proficient in English and had normal or corrected-to-normal vision. Participants ranged in age from 20 to 73 years ($M = 37.93$, $SD = 11.48$); 105 identified as male (52.5\%), 94 as female (47.0\%), and one preferred not to disclose their gender (0.5\%). Each participant received USD 8 in compensation, and the study took approximately 45 minutes to complete on average. Each participant was assigned to one of the 20 task sets and completed eight drawing tasks in randomized order.
Submitted drawings were manually reviewed by a researcher, and five participants were excluded because all of their drawings consisted of random scribbles indicating clear non-engagement with the task.
Following these exclusions, additional participants were recruited until 10 valid participants had completed each task set, resulting in 200 valid participants and 1,600 human-drawn annotation sketches, with 10 independently drawn annotations for each chart-caption pair.

We reused the drawing interface from the first study but replaced the previous evaluation questions with a new set of post-task questions.
For each task, participants read the caption and drew visual highlights directly on the corresponding chart to support caption understanding.
After completing their drawing, participants rated on seven-point scales, ranging from \emph{very easy} to \emph{very difficult}, how difficult it was to understand the chart, understand the caption, and connect the caption to the relevant parts of the chart.
They then selected one or more intents for their visual highlights from \emph{Identify}, \emph{Compare}, \emph{Summarize}, \emph{Present}, and \emph{Other}~\cite{rahman2024qualitative}.
To help interpret the drawings, participants also provided an open-ended
explanation of what they intended to convey, since free-form sketches can
leave their intended meaning ambiguous~\cite{li2026sketchdynamics}.
Finally, they described what was missing or unclear in their drawing and what they would add or change to improve it.

\subsection{Metadata Schema}
\label{sec:dataset:metadata_schema}

Collections of visual artifacts can be supplemented with structured descriptions of properties that are not explicitly represented in the raw artifacts.
Prior visualization datasets have taken this approach by pairing collected visual artifacts with structured annotations based on labeling schemes developed by researchers or visualization practitioners~\cite{deng2022visimages, chen2025visanatomy}.
Following this approach, we add structured metadata to collected sketches to describe how each sketch relates to its caption and chart and how individual marks serve different communicative functions.
Our metadata schema comprises three top-level dimensions: \textit{Caption Coverage} and \textit{Caption Extension} at the sketch level, and \textit{Annotation Form \& Role} at the mark level. Prior work suggests that readers form takeaways by integrating information from both charts and captions and may draw on visual patterns or contextual knowledge beyond what is explicitly stated in the caption~\cite{kim2021towards,lundgard2021accessible}. Accordingly, the sketch-level dimensions capture both the extent to which sketches reflect the captioned message and whether they incorporate information beyond it.
To derive the metadata schema, two researchers independently open-coded a randomly selected subset of 160 sketches together with the corresponding intent responses, then compared and refined the codes through discussion.
Redundant, overly specific, or semantically overlapping codes were consolidated into broader categories that could be applied consistently across sketches.

\par\vspace{0.3\baselineskip}
\noindent\textbf{Caption Coverage} characterizes how much of the information conveyed in the caption is represented in the sketch.
We distinguish three categories: \textit{Entire}, \textit{Portion}, and \textit{None}.
\textit{Entire} indicates that the sketch represents all major information, relationships, or patterns described in the caption.
\textit{Portion} indicates that the sketch represents at least one meaningful part of the caption but does not cover all of its information.
\textit{None} is assigned when the sketch does not represent information described in the caption, such as when participants trace or recolor chart elements without expressing the information described in the caption.

\par\vspace{0.3\baselineskip}
\noindent\textbf{Caption Extension} characterizes information expressed in the sketch that is not explicitly stated in the corresponding caption.
We distinguish two categories based on the source of that information: \textit{Chart-grounded} and \textit{External source}.
\textit{Chart-grounded} refers to additional information that can be directly derived or reasonably inferred from the chart, such as an exact numeric value, an additional category, or a visible trend.
\textit{External source} refers to information that is not supported by the chart or caption and relies on external knowledge, assumptions, or participant-defined interpretations.
For example, a participant may assign a custom or domain-specific meaning to a symbol not present in the chart.

\begin{table}[t]
\centering
\caption{\textit{Annotation Forms \& Roles} used for sketch labeling, with a brief description of what each Role communicates or visually indicates.}
\label{tab:metadata_schema}
% \vspace{-4pt}

\small
\renewcommand{\arraystretch}{1}
\setlength{\tabcolsep}{3pt}

\begin{tabularx}{\columnwidth}{
    @{}>{\raggedright\arraybackslash\leftskip=0.0em}p{0.37\columnwidth}
    >{\raggedright\arraybackslash}p{0.61\columnwidth}
}
\toprule
\multicolumn{1}{@{}l}{\textbf{Form / Role}} & \textbf{Function} \\
\midrule

\multicolumn{1}{@{}l}{\textbf{\textit{Text}}} & \\
\textit{T1.} Show Value & An exact or approximate numeric value \\
\textit{T2.} Explain & Clarification or summary of caption or annotation \\
\textit{T3.} Show Rank & Relative order or rank of data elements \\
\textit{T4.} Label & Identity of a data element, category, or annotation \\

\addlinespace[2pt]
\multicolumn{1}{@{}l}{\textbf{\textit{Line}}} & \\
\textit{L1.} Highlight & Additional cue for a specific element or region \\
\textit{L2.} Indicate Value & A specific data value or position \\
\textit{L3.} Connect Elements & Correspondence or relation between elements \\
\textit{L4.} Differentiate Categories \& Annotations & Distinctions among categories, groups, or annotations \\
\textit{L5.} Show Trend / Distribution & An overall trend or distribution \\
\textit{L6.} Show Extent & Distance or span between values, positions, or elements \\
\textit{L7.} Show Boundary & A baseline, threshold, divider, or boundary \\

\addlinespace[2pt]
\multicolumn{1}{@{}l}{\textbf{\textit{Color}}} & \\
\textit{C1.} Differentiate Annotations & Distinctions among participant-added annotations \\
\textit{C2.} Differentiate Chart Elements & Distinctions among chart or data elements \\

\addlinespace[2pt]
\multicolumn{1}{@{}l}{\textbf{\textit{Enclosure}}} & \\
\textit{E1.} Show Trend / Distribution & An overall trend or distribution \\
\textit{E2.} Highlight & Additional cue for a specific element or region \\
\textit{E3.} Group Elements & A set of elements interpreted together \\

\addlinespace[2pt]
\multicolumn{1}{@{}l}{\textbf{\textit{Glyph}}} & \\
\textit{G1.} Show Change & A change in value, direction, or trend \\
\textit{G2.} Highlight & Additional cue for a specific element or region \\
\textit{G3.} Symbolize & A participant-defined semantic meaning \\
\textit{G4.} Show Relation & Relative relationship between data elements \\
\textit{G5.} Indicate Exclusion & Elements or regions to exclude or ignore \\
\textit{G6.} Indicate Polarity & Positive/negative or favorable/unfavorable meaning \\

\bottomrule
\end{tabularx}
\end{table}

\begin{figure}[t]
\centering
\includegraphics[width=0.98\columnwidth]{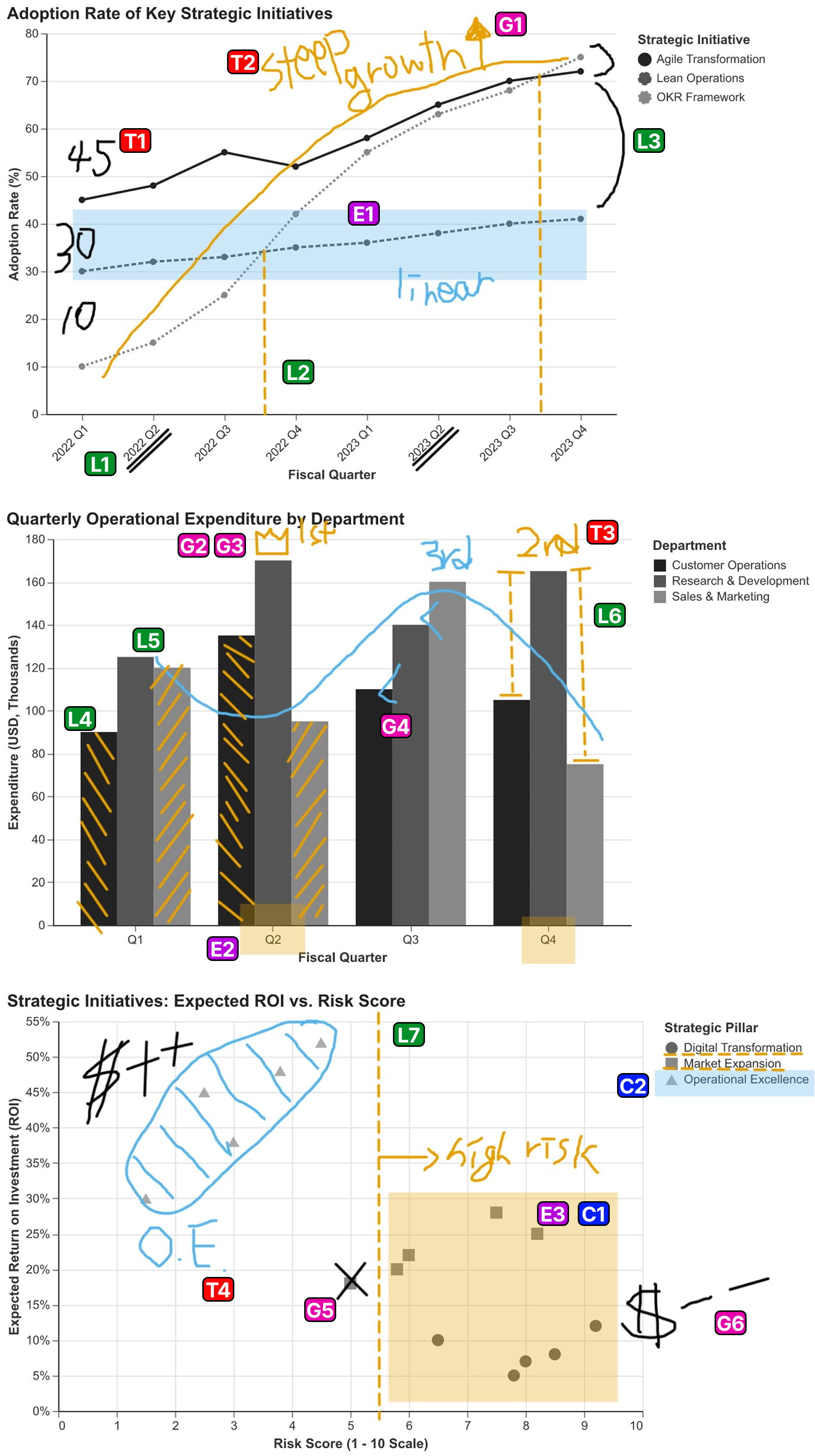}
\caption{
Illustrative Examples of Annotation Form-Role combinations. 
Codes correspond to the Forms and Roles in Table~\ref{tab:metadata_schema}.}
\label{fig:metadata_examples}
\end{figure}

\begin{figure*}[t]
    \centering
    \includegraphics[width=\textwidth]{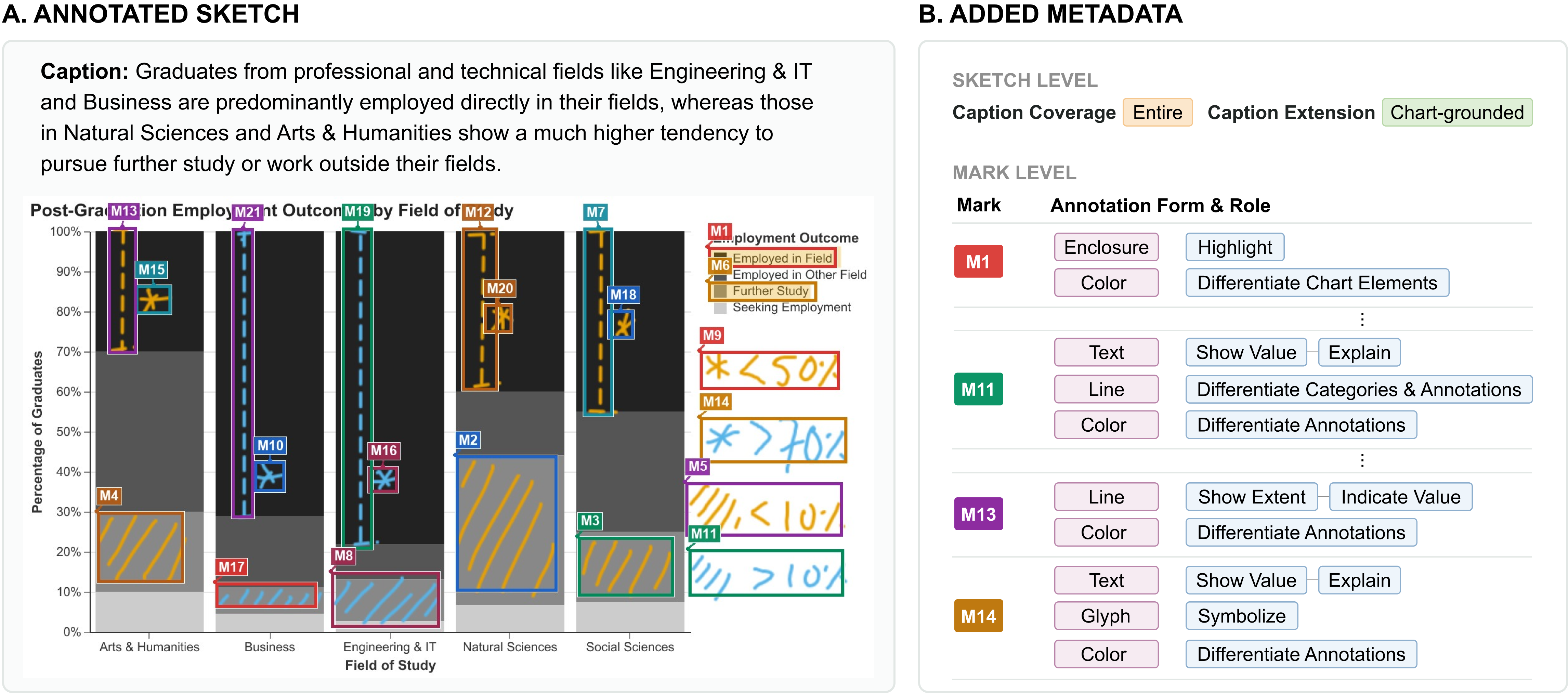}
    \caption{Example of an annotation sketch and its associated metadata. 
    (A) An annotation sketch is shown together with the corresponding caption. 
    Each sketch mark is identified with a bounding box and a unique mark ID. 
    (B) The dataset attaches structured metadata at both the sketch and mark levels, including \textit{Caption Coverage} and \textit{Caption Extension} for the overall sketch, and \textit{Annotation Form} and \textit{Role} for each identified mark. 
    A single sketch mark may receive multiple Form labels and multiple associated Roles when supported by the annotation.}   
    \label{fig:metadata_labeling.png}
    % \vspace{-8pt}
\end{figure*}

\par\vspace{0.3\baselineskip}
\noindent\textbf{Annotation Form \& Role}
characterize the visual properties and communicative functions of individual sketch marks.
\textit{Annotation Form} describes the visual form of a mark, whereas \textit{Annotation Role} describes the communicative function it serves.
For \textit{Annotation Form}, we adapted the taxonomy of annotation types proposed by Rahman et al.~\cite{rahman2024qualitative} as an initial framework.
Their taxonomy defines seven annotation types: Enclosure, Connector, Text, Glyph, Color, Indicator, and Geometric.
We excluded Geometric because no corresponding instances were observed in our sketches.
We also combined Connector and Indicator, which share a line-based appearance, into a single \textit{Line} category, yielding five Forms: Enclosure, Line, Text, Glyph, and Color.
A single mark may receive multiple Form labels and multiple Role labels when applicable.
Table~\ref{tab:metadata_schema} summarizes the Roles associated with each Form and the communicative function of each Role.
We illustrate these Form-Role combinations below using the examples in Fig.~\ref{fig:metadata_examples}.

\textbf{Text} denotes textual annotations, including numbers, words, phrases, labels, and abbreviations.
In Fig.~\ref{fig:metadata_examples}, \textit{Show Value}~(T1) writes the numeric value ``45,'' whereas \textit{Explain}~(T2) adds the interpretive phrase ``steep growth'' rather than a value or identifier.
\textit{Show Rank}~(T3) uses ordinal labels such as ``1st,'' ``2nd,'' and ``3rd'' to encode the ordering of data elements.
\textit{Label}~(T4), in contrast, names a marked element or region, as illustrated by ``O.E.,'' an abbreviation for Operational Excellence.

\textbf{Line} refers to line-based annotations used to point to, connect, trace, span, or separate chart elements and regions.
\textit{Highlight}~(L1) appears as short underline marks beneath selected axis labels, without encoding an additional value or relationship.
\textit{Indicate Value}~(L2) uses a dashed vertical line to mark a particular position on the x-axis, whereas \textit{Connect Elements}~(L3) connects the three endpoints of the plotted lines.
\textit{Differentiate Categories \& Annotations}~(L4) uses repeated hatching over selected bars to distinguish them from the other bars.
\textit{Show Trend / Distribution}~(L5) traces the distribution of the rightmost bars across the four groups.
\textit{Show Extent}~(L6) uses a dashed line to show the height difference between two adjacent bars.
\textit{Show Boundary}~(L7) uses a vertical dashed line to divide the scatterplot into two regions.

\textbf{Color} captures cases in which hue itself distinguishes chart elements or participant-added annotations.
\textit{Differentiate Annotations}~(C1) uses blue for the annotations grouping the upper-left points and yellow for those grouping the lower-right points.
\textit{Differentiate Chart Elements}~(C2) uses blue to distinguish a legend entry.

\textbf{Enclosure} refers to closed shapes that surround chart elements, annotations, or regions.
\textit{Show Trend / Distribution}~(E1) uses a blue enclosure spanning several points of the lower line to capture its relatively flat pattern.
\textit{Highlight}~(E2) encloses the Q2 axis label to draw attention to that part of the chart.
\textit{Group Elements}~(E3) uses a yellow enclosure to group the lower-right scatterplot points together.

\textbf{Glyph} refers to semantic symbols, such as short arrows, icons, and other symbolic marks that convey meaning.
\textit{Show Change}~(G1) uses an upward arrow at the end of the ``steep growth'' annotation to indicate an increase.
\textit{Highlight}~(G2) places a crown-like symbol above the tallest bar to draw attention to it.
\textit{Symbolize}~(G3) uses the same crown-like symbol to represent the top-ranked bar.
\textit{Show Relation}~(G4) uses an inequality symbol between bars to visually relate their values.
\textit{Indicate Exclusion}~(G5) places an ``X'' next to a scatterplot point to mark it as excluded.
\textit{Indicate Polarity}~(G6) contrasts ``$++$'' and ``$--$'' annotations to indicate positive and negative interpretations.

\subsection{LLM-Assisted Labeling and Validation}
\label{sec:dataset:labeling}
Recent HCI work~\cite{ma2026negotiating,shi2026aligning} has used LLMs in tasks that require interpretive judgments, with model outputs compared against human judgments.
Within visualization research, Hines and Ottley~\cite{hines2026charting} developed a coding scheme through human analysis, used an LLM to apply it, and validated the resulting labels against human coding.
Given that several of our metadata labels also require interpretation, we involved human researchers in both schema development and validation.

Each of the 1,600 sketches was then labeled independently by three models: Gemini 3.7 Flash, Claude Fable 5, and GPT-5.6 Sol. 
Each model received the original chart, the participant's annotated sketch, the corresponding caption, the participant's open-ended intent response, and the finalized metadata schema and labeling rules.

First, the models identified individual sketch marks in each sketch by comparing the annotated sketch with the original chart. 
Each identified mark was then assigned all applicable \textit{Annotation Form \& Role} labels. 
At the sketch level, the models additionally labeled \textit{Caption Coverage} and \textit{Caption Extension}. 
The participant's open-ended intent response was used as supporting evidence for interpreting visually present annotations. 
Before returning the results in structured JSON format, each model performed a final consistency check to verify compliance with the metadata schema and labeling rules.

Following independent labeling, we consolidated the outputs across the three models. 
We first performed mark clustering to group detections referring to the same visual marks.
Marks were matched based on spatial overlap using an intersection-over-union threshold of $\mathrm{IoU}\geq0.35$ together with an area-ratio threshold of at least $0.25$. 
A cluster was retained when supported by at least two models, and its final bounding box was defined using the coordinate-wise median of the matched bounding boxes.

For each consensus mark, \textit{Annotation Form \& Role} labels were determined via a majority vote, retaining labels agreed upon by at least two models~\cite{byun2026cradle}. 
Seven cases lacking a valid majority were manually reviewed and labeled by the authors based on model outputs and labeling rules. 
We applied the same majority-consensus rule to sketch-level labels (\textit{Caption Coverage} and \textit{Caption Extension}). 
Cases in which the models produced mutually conflicting outputs were manually adjudicated by the researchers (six cases for \textit{Caption Coverage} and seven for \textit{Caption Extension}). 
Detailed prompt templates and implementation specifics are provided in the supplementary material.

To validate the LLM-generated labels, we randomly selected one participant from each of the 20 task sets, yielding 160 sketches (8 per participant). This sampling yielded one human-annotated sketch for each of the 160 chart-caption pairs. An author independently annotated these sketches using the same metadata schema and coding rules as the LLM pipeline. We measured human-LLM agreement on five aspects: \textit{Caption Coverage}, \textit{Caption Extension}, mark identification, \textit{Annotation Form}, and \textit{Annotation Role}. \textit{Caption Coverage} and \textit{Caption Extension} were evaluated at the sketch level using exact-match agreement, with Extension requiring an exact match of the \textit{Caption Extension} label set. Mark identification matched human and LLM marks by spatial overlap (IoU $\geq$ 0.35) and was summarized with an F1 score. On matched marks, \textit{Annotation Form} required an exact match of the Form set. \textit{Annotation Role} was evaluated only when Form agreed; Form-mismatched marks were excluded from the Role score. Mark identification, Form, and Role were micro-averaged across the sample.

The results showed 92.5\% agreement for \textit{Caption Coverage}, 83.8\% for \textit{Caption Extension}, an F1 score of 95.2\% for mark identification, 85.3\% for \textit{Annotation Form}, and 82.8\% for \textit{Annotation Role}. 

\subsection{Annotation Patterns} 
\label{sec:dataset:annotation_patterns}

Participants' sketches demonstrated a strong grounding in the corresponding chart-caption pairs. Among the 1,600 sketches, 52.2\% represented the \textit{Entire} caption, while 43.4\% captured a \textit{Portion}, indicating that 95.6\% of all sketches incorporated at least some caption information.  At the same time, participants did not necessarily restrict their annotations to what was explicitly stated in the caption. Beyond-caption annotation appeared in 523 sketches (32.7\%). These additions were overwhelmingly \textit{Chart-grounded} (30.9\% of all sketches). In contrast, only 1.5\% contained exclusively \textit{External source} information, and 0.3\% included both. These findings reveal that even when participants extended their annotations beyond the provided caption, they primarily elaborated on values, patterns, or categories already present in the chart rather than introducing external knowledge.

At the mark level, \textit{Color} was the most prevalent Form (38.0\% of all forms), primarily due to its frequent co-occurrence with other visual forms. Specifically, the combination of \textit{Color + Enclosure} appeared in 2,486 marks (34.1\%), and \textit{Color + Line} accounted for 1,381 marks (18.9\%). Together, these two compound forms constituted more than half (53.0\%) of all consensus marks, highlighting the primary role of color as a complementary visual form.

Visual Forms exhibited distinct communicative functions, although these associations were not exclusive. Primary Roles aligned strongly with specific Forms: \textit{Enclosure} overwhelmingly served to \textit{Highlight} (98.2\%), \textit{Line} most frequently functioned to \textit{Show Trend / Distribution} (55.7\%), \textit{Text} primarily aimed to \textit{Show Value} (53.9\%), \textit{Glyph} served to \textit{Show Change} (43.3\%), and \textit{Color} was mainly utilized to \textit{Differentiate Chart Elements} (62.0\%). However, each Form extended beyond its dominant function. For instance, Enclosures frequently served to \textit{Group Elements} (26.0\%); Lines additionally supported highlighting, value indication, connection, and range expression; Text regularly provided explanatory keywords (30.5\%) or labels (22.3\%); and Glyphs conveyed semantic concepts, polarity, and relationships. Collectively, these patterns demonstrate that while specific visual Forms carried signature functional roles, participants flexibly utilized the same Form to fulfill multiple communicative functions based on contextual needs.

\section{Discussion}
\label{sec:discussion}

\subsection{Sketches as Guidance for Chart Annotation Generation}
Our first study found that the benefit of human sketch input varied across chart–caption conditions.
Specifically, sketch-conditioned annotations showed a measurable advantage only for Complex–L3 pairs, where baseline annotations were rated least helpful.
One possible explanation is that these pairs leave more uncertainty about which chart elements and relationships should be emphasized: complex charts contain more candidate marks and series, while L3 captions often describe patterns spanning multiple elements or regions.
In such cases, a rough sketch serves to indicate which parts of the chart are most relevant to the intended message.

Our qualitative results further showed that 90.6\% of preferred sketch-conditioned outputs reflected participants' sketches, while the majority also introduced additional annotation forms that were not present in the original sketches.
Notably, even when the baseline output was preferred, the sketch-conditioned output still reflected the participant's sketch in 76.3\% of cases. 
Together, these findings suggest that the value of sketch conditioning may lie not simply in reproducing what users draw, but in maintaining visual correspondence with their sketches while adding additional visual cues that are not present in the sketch.
However, participants' explanations also showed that sketch input did not eliminate problems in the generated annotations. 
When sketch-conditioned outputs were not preferred, participants commonly cited inaccurate generation or alterations to the original chart, while also noting visual clutter.

Taken together, these results suggest that sketches are particularly useful when there are many plausible candidates for annotation or when users want the generated output to reflect a specific annotation intent. In such cases, the sketch can narrow the set of relevant chart elements while still allowing the model to draw on the surrounding context and introduce additional annotation forms where appropriate.

\subsection{Empirical Resource for Intent-Aware Annotation}
\dsname{} serves as an empirical resource for studying intent-aware chart annotation by combining hand-drawn sketches with participants' intent descriptions, self-reported expressive limitations, and structured metadata. 
Together, these elements connect participants' rough visual marks with their stated intent and with structured descriptions of how those marks relate to the chart and caption.

The annotation patterns illustrate the value of capturing these different aspects together. 
Participants' sketches were strongly grounded in the corresponding chart–caption pairs, but a substantial subset also included additional chart-grounded information beyond what was explicitly stated in the caption. 
Visual Forms also tended to serve characteristic Roles while supporting multiple communicative functions.
These patterns suggest that caption-supporting annotations cannot always be understood as direct visual restatements of caption content, and that the meaning of a rough mark may depend on its chart–caption context as well as its visual form.

\dsname{} can support the design and evaluation of intent-aware annotation systems.
Its structured metadata can help characterize how communicative roles are expressed through different visual forms and how annotations extend caption content, providing empirical grounding for annotation grammars and authoring interfaces.
The paired sketches and intent descriptions can support evaluations of whether systems interpret rough marks in ways consistent with participants’ stated intent.
Prior work has shown that explicit communicative goals can influence visualization design choices~\cite{leerobbins2021learning}, highlighting the value of retaining participants' stated intent alongside their sketches.
The self-reported expressive limitations provide material for studying where sketch input is insufficient and what additional authoring support may be needed.
Multiple independent sketches for each chart-caption pair further provide a basis for studying variation in annotation strategies and for exploring interfaces that accommodate multiple plausible visual expressions.

\subsection{Limitations and Future Work}
Our study has several limitations that suggest directions for future work. 
First, our current corpus was not designed to systematically cover different data fact types and levels of perceptual complexity. 
Different data fact types may lead to different L3 content, while properties such as the number and density of marks may affect how difficult those patterns are to perceive and relate to the caption. 
Future work could construct stimuli using existing data-fact taxonomies while systematically varying such visual properties, enabling a more structured investigation of how semantic and perceptual factors shape annotation strategies and the value of sketch input.

Second, our evaluation used a single MLLM.
The observed benefit of sketch input may therefore depend on the capabilities of the model used in our study.
Future work could test whether the same pattern holds across different MLLMs.

Third, we treated sketch guidance as a one-shot input.
Participants provided their sketches before generation and evaluated the resulting outputs without opportunities for further interaction.
Recent AI-assisted visualization systems have supported iterative authoring by combining multiple forms of user input and allowing users to refine previous results through structured interactions or direct manipulation~\cite{wang2025dataformulator2,wen2026multimodal}.
Future work could examine similar workflows for chart annotation, in which users revise their sketches, directly edit generated annotations, or provide additional feedback after viewing each result.

Finally, our study does not resolve how a model should interpret ambiguity in rough sketch input.
Spatial or graphical imprecision may reflect drawing error, rough execution, or intentional abstraction, making it difficult to determine which aspects of a sketch should be preserved.
Future systems must therefore balance correcting inconsistencies with the chart or data against retaining visual attributes that represent meaningful user guidance.
Future work could explore uncertainty-aware or interactive mechanisms that allow users to clarify ambiguous marks rather than having the model either reproduce or override them.
\section{Conclusion}
\label{sec:conclusion}

We investigated when rough human sketches can support MLLM chart annotation generation and what information users communicate through them. 
Sketch conditioning provided a clear advantage specifically for complex charts paired with L3 captions, where ambiguity exists about which visual elements and relationships to emphasize. 
This suggests that rough sketches can effectively convey annotation intent while allowing the model to adaptively add context-appropriate visual annotations.
Building on this result, we introduced AnnoSketch, a dataset of 1,600 human-drawn annotation sketches across 160 Complex–L3 chart–caption pairs, paired with participants’ stated intent, self-reported limitations, perceived difficulty, and structured sketch- and mark-level metadata. 
Our analysis shows that sketches are strongly grounded in caption content while often extending it with chart-grounded information, and that the same visual form can serve different communicative roles depending on context.
Together, these findings provide a foundation for intent-aware chart annotation systems that selectively solicit human guidance, interpret rough marks in context, and translate them into clear, polished visual annotations.

\bibliographystyle{abbrv-doi-hyperref}
\bibliography{Ref}

\appendix\newpage
\section{Appendices}
\label{sec:appendices_inst}

\subsection{Generation Prompts}
\label{app:generation-prompts}

This section presents condensed versions of the prompts and
multimodal inputs used for chart, caption, and baseline annotation
generation. 
Headings enclosed by equal signs~(===) indicate API message roles. 
Angle brackets~(<>) identify prompt sections and curly-braced labels~(\{\}) indicate values or images dynamically inserted for each request. Full prompt is available in the \href{https://osf.io/ad5r6/overview?view_only=791aee36e8b54cafa6a57c1cc25b6438}{OSF repository}.

%%%%%%%%%%%%%%%%%%%%%%%%%%%%%%%%%%%%%%%%%%%%%%%%%%%%%%%%%%%%
\subsubsection{Chart and Caption Generation}
\label{app:chartcaption-generation-prompts}

\begin{lstlisting}[
  style=judgeprompt,
  title={Prompts for Chart Generation}
]
=== SYSTEM PROMPT ===
You are a chart generation expert. Please write complete Vega-Lite v5 chart specifications as JSON.
During the generation process, please follow the style rules and quality rules strictly.

Input: chart subtype + topic. Invent realistic fictional data in data.values.

<CHART APPEARANCE, ENCODING, AND LAYOUT RULES>

<DATA DIVERSITY, REALISM, AND READABILITY RULES>

Return one valid Vega-Lite JSON object containing $schema, data.values, title, mark or layer, encoding, and config.

=== USER PROMPT ===
Create one chart.
Chart subtype: {CHART_TYPE} Topic: {TOPIC}
Use a specific scenario within this topic. Match the chart subtype exactly.
\end{lstlisting}

\begin{lstlisting}[
  style=judgeprompt,
  title={Prompts for L2 and L3 Caption Generation}
]
=== SYSTEM PROMPT ===
You are a data visualization expert. Your task is to write two types of captions for a chart.

You receive:
1) a chart image,
2) chart metadata and underlying data values as ground truth.

Write exactly TWO captions for the chart.
In this process, please consider the following instructions for each caption type.

[Instruction for Caption L2]
- Examine the chart image and generate one Level 2 (Statistical & Relational) caption.
- The caption should describe objective statistical facts or relationships that can be directly inferred from the visible chart.
- Combine at least two types of Level 2 information, such as values, rankings, extrema, differences, correlations, or comparisons.

<L2 CONTENT SCOPE, NUMERIC PRECISION, AND EXCLUSION RULES>

[Instruction for Caption L3]
- Examine the chart image and generate one Level 3 (Perceptual & Cognitive) caption.
- The caption should describe higher-level patterns, trends, structures, exceptions, or overall insights visible in the chart.
- Synthesize multiple visible marks, groups, or trends into one natural-language takeaway.

<L3 PATTERN SYNTHESIS, INTERPRETATION, AND EXCLUSION RULES>

<CAPTION LENGTH, GROUNDING, NON-HALLUCINATION, AND TONE RULES>

[JSON output]

{RESPONSE FIELDS AND CAPTION-TYPE CONSTRAINTS}

Return valid JSON matching the required schema.

=== USER PROMPT ===
{RENDERED_CHART_IMAGE}

Generate two captions for this chart. Follow the system instructions for L2 and L3.

Chart subtype: {CHART_TYPE} Topic: {TOPIC} Chart ID: {CHART_ID}

## Chart metadata
Title: {CHART_TITLE} X-axis: {X_AXIS_TITLE} Y-axis: {Y_AXIS_TITLE}

## Underlying data values
{DATA_VALUES_AS_JSON}

Output:
- caption_type L2 -- Level 2 (Statistical & Relational)
- caption_type L3 -- Level 3 (Perceptual & Cognitive)
- Each caption: caption_type and text only
\end{lstlisting}

%%%%%%%%%%%%%%%%%%%%%%%%%%%%%%%%%%%%%%%%%%%%%%%%%%%%%%%%%%%%
\subsubsection{Annotation Generation}
\label{app:baseline-annotation-prompt}
Both annotation conditions used Gemini 3.1 Flash, with no explicit temperature or other sampling-parameter configuration.

\begin{lstlisting}[
  style=judgeprompt,
  title={Prompt for Baseline Annotation Generation}
]
=== USER PROMPT ===
{ORIGINAL_CHART_IMAGE}

**Information:**
Caption: {CAPTION}

You will receive one image:
- **Original chart:** the base chart with no annotations. This is the fixed chart layer.

**Task:**
1. Analyze the caption together with the original chart. Identify both:
   - **What to express** -- which data, regions, trends, comparisons, or relationships the caption emphasizes and that are supported by the chart.
   - **How to express it** -- suitable visual emphasis that connects the chart to the caption.
2. Draw visual annotations on top of the original chart based on the identified "What" and "How."

<CHART PRESERVATION, ANNOTATION LEGIBILITY, TEXT, AND STYLING RULES>
\end{lstlisting}

\begin{lstlisting}[
  style=judgeprompt,
  title={Prompt for Sketch-Conditioned Annotation Generation}
]
=== USER PROMPT ===
**Information:**
Caption: {CAPTION}

You will receive two images in order:
- **Image 1 -- Original chart:** the base chart with no participant highlights. This is the fixed chart layer.
- **Image 2 -- Participant drawing:** the same chart with the participant's rough visual highlights on top.

**Task:**
1. Compare Image 1 and Image 2. From the participant's drawing, identify both:
   - **What they are trying to express** -- which data, regions, trends, comparisons, or relationships their highlights target, and how those highlights connect to the caption.
   - **How they are trying to express it** -- the communicative role of their marks, such as emphasis, grouping, direction, or contrast.
2. Draw visual annotations on top of Image 1 based on the identified "What" and "How."

<PARTICIPANT-DRAWING INTERPRETATION AND REFINEMENT RULES>

<CHART PRESERVATION, ANNOTATION LEGIBILITY, TEXT, AND STYLING RULES>

Image 1 -- Original chart (fixed base; do not alter underlying chart pixels):
{ORIGINAL_CHART_IMAGE}

Image 2 -- Participant drawing (chart + rough highlights; infer what they express and how they emphasize it -- do not copy its pixels):
{PARTICIPANT_DRAWING_IMAGE}
\end{lstlisting}

%%%%%%%%%%%%%%%%%%%%%%%%%%%%%%%%%%%%%%%%%%%%%%%%%%%%%%%%%%%%
\subsection{LLM-Assisted Metadata Labeling}
\label{app:metadata-labeling}

\subsubsection{Prompt for Metadata Labeling}
\label{app:annotation-coding-rules}

The same labeling prompt was used for every labeling trial.
The model was instructed to first identify participant-added marks by
comparing the original charts and sketches, and then assign sketch-level
and mark-level metadata in a fixed order.
Full prompt is available in the \href{https://osf.io/ad5r6/overview?view_only=791aee36e8b54cafa6a57c1cc25b6438}{OSF repository}.

\begin{lstlisting}[
  style=judgeprompt
]
=== SYSTEM PROMPT ===
You are an expert annotation coder for chart sketch labeling.

You must follow the Annotation Coding Rules below EXACTLY and in order.
You must use the Metadata Schema below as the PRIMARY coding criteria.
Use only exact label strings from the Metadata Schema.

=== CRITICAL OUTPUT / FORM RULES ===

1) Inventory annotations by physical mark.
   Assign mark_id = M1, M2, M3, ...
   Each output entry has exactly one Form.

   If one physical mark has multiple Forms, create
   separate entries sharing the same mark_id:
   M1-1, M1-2, ...

2) Short symbolic arrows -> Glyph.
   Extended arrowed paths used as trends, guides,
   or connectors -> Line.

3) Closed or nearly closed boxes/circles -> Enclosure.

   Color must not stand alone; it must co-occur
   with another Form on the same physical mark.

4) Use exact Metadata Schema strings for all
   Caption Coverage, Caption Extension,
   Annotation Form, and Annotation Role labels.


=== ANNOTATION CODING PROCEDURE ===

1. Identify Participant-added Annotations
   - Compare the original chart with the participant sketch.
   - Record every visible participant-added physical mark.
   - Ignore chart elements already present in the original image.
   - Treat participant intent as supporting evidence only
     for annotations that are visually present.

2. Determine Caption Coverage
   - Determine whether the sketch represents the Entire caption,
     a Portion of the caption, or None of the caption information.
   - Select exactly one label.

3. Determine Caption Extension
   - Identify information expressed in the sketch that is not
     explicitly stated in the caption.
   - Label additional information as Chart-grounded and/or
     External source when applicable.

4. Identify Annotation Form
   - Assign all applicable Forms to each physical mark.
   - If one physical mark has multiple Forms, create separate
     Form entries sharing the same mark_id.
   - Color cannot be assigned as a standalone Form.

5. Identify Annotation Role
   - Assign all Roles supported by each Form.
   - Roles are not mutually exclusive.
   - Use only Roles defined for the assigned Form.

6. Perform Final Validation
   - Verify that every visible participant-added mark is represented.
   - Verify Form--Role compatibility.
   - Verify exact Metadata Schema label strings.
   - Verify mark_id and annotation_id consistency.
   - Verify that every assigned Role includes an
     evidence-based reason.

7. Return the Final JSON Output.


<METADATA SCHEMA DEFINITIONS>


=== OUTPUT REQUIREMENTS ===

Return ONLY one valid JSON object matching the required
structured output schema.

Do not include explanatory prose before or after the JSON.

Every reason must be concise and evidence-based.

Do not invent annotations mentioned in participant intent
but not visible in the participant sketch.


=== USER PROMPT ===

Label this chart annotation trial.

You will receive two images in order:
1) Original chart image (no participant annotations)
2) Annotated chart image (participant sketch)

**Information:**
Caption: {CAPTION}

Participant free-form intent
(supporting evidence only; do not invent annotations
that are not visible):
{PARTICIPANT_INTENT}

Follow the required coding order:
1. Identify Participant-added Annotations
2. Determine Caption Coverage
3. Determine Caption Extension
4. Identify Annotation Form
5. Identify Annotation Role
6. Perform Final Validation
7. Return the Final JSON Output

Checks before finalizing:
- Short symbolic arrows -> Glyph; other arrows -> Line.
- Closed boxes/circles -> Enclosure.
- Color must co-occur with another Form on the same mark.
- mark_id = M1/M2/...; annotation_id = M1-1/M1-2/...
- Roles must be valid for each Form.

Image 1 -- Original chart:
{ORIGINAL_CHART_IMAGE}

Image 2 -- Participant sketch:
{PARTICIPANT_SKETCH_IMAGE}
\end{lstlisting}

\end{document}